\documentclass[10pt,conference]{IEEEtran}
\IEEEoverridecommandlockouts

\usepackage{tikz}
\usepackage{amsmath}
\usepackage{url}
\usepackage{balance}
\usepackage{listings}
\usepackage{calc}
\usepackage{bbding}
\usepackage{pifont}
\usepackage{wasysym}
\usepackage{xcolor}
\usepackage{makecell}
\usepackage{graphicx}
\usepackage{caption,subcaption}
\usepackage{enumitem}
\usepackage{supertabular}
\usepackage{colortbl}
\usepackage{microtype}
\usepackage{hyperref}
\usepackage{fancyvrb}
\usepackage{multicol}
\usepackage{rotating}
\usepackage{enumitem}
\usepackage{algorithmic}
\usepackage{xspace}
\usepackage{braket}
\usepackage{booktabs}
\usepackage[most]{tcolorbox}
\usepackage{minted}
\tcbuselibrary{minted}

\usepackage[ruled,vlined,linesnumbered]{algorithm2e}
\usepackage{tikz}
\usepackage{mdframed}
\usepackage{multirow}
\usepackage{bigstrut}
\usepackage{mathtools}
\usepackage{mathpartir}
\usepackage{comment}
\usepackage{pifont}

\definecolor{myblue}{RGB}{0,32,96}

\newcolumntype{P}[1]{>{\raggedright\arraybackslash}p{#1}}

\definecolor{todocolor}{rgb}{0.9,0.1,0.1}
\definecolor{ylcolor}{rgb}{0.7,0.7,0.3}

\begin{document}
\bstctlcite{IEEEexample:BSTcontrol}

\title{Is Measurement Enough? Rethinking Output Validation in Quantum Program Testing}

\DeclareRobustCommand*{\IEEEauthorrefmark}[1]{%
  \raisebox{0pt}[0pt][0pt]{\textsuperscript{\footnotesize #1}}%
}

\author{
    \IEEEauthorblockN{
        Jiaming Ye\IEEEauthorrefmark{1},
        Xiongfei Wu\IEEEauthorrefmark{2},
        Shangzhou Xia\IEEEauthorrefmark{3},
        Fuyuan Zhang\IEEEauthorrefmark{4}, 
        Jianjun Zhao\IEEEauthorrefmark{3}\thanks{Fuyuan Zhang is with the State Key Laboratory of Blockchain and Data Security, Zhejiang University, Hangzhou, China. This work was supported in part by JSPS KAKENHI Grant No. JP24K14908, and in part by JST SPRING Grant No. JPMJSP2136. }
        }
    \IEEEauthorblockA{
        \IEEEauthorrefmark{1}Southwest Jiaotong University, China
        \IEEEauthorrefmark{2}University of Luxembourg, Luxembourg 
        \IEEEauthorrefmark{3}Kyushu University, Japan \\
        \IEEEauthorrefmark{4}Zhejiang University, China
    }
}


\maketitle

\begin{abstract}

As quantum computing continues to emerge, ensuring the quality of quantum programs has become increasingly critical. Quantum program testing has emerged as a prominent research area within the scope of quantum software engineering. While numerous approaches have been proposed to address quantum program quality assurance, our analysis reveals that most existing methods rely on measurement-based validation in practice. However, due to the inherently probabilistic nature of quantum programs, measurement-based validation methods face significant limitations. 

To investigate these limitations, we conducted an empirical study of recent research on quantum program testing, analyzing measurement-based validation methods in the literature. Our analysis categorizes existing measurement-based validation methods into two groups: distribution-level validation and output-value-level validation. We then compare measurement-based validation with statevector-based validation methods to evaluate their pros and cons. Our findings demonstrate that measurement-based validation is suitable for straightforward assessments, such as verifying the existence of specific output values, while statevector-based validation proves more effective for complicated tasks such as assessing the program behaviors.

\end{abstract}


\section{Introduction}


The growing applications of quantum software engineering have increased the need for quality assurance of quantum programs, leading researchers to propose various testing approaches~\cite{miranskyy2019testing,intergration-testing-01}, such as search-based testing~\cite{search-based-01, search-based-02}, combinatorial testing~\cite{combinatorial-01,combinatorial-02}, property-based testing~\cite{property-based-testing-01}, concolic testing~\cite{concolic-testing-01}, and metamorphic testing~\cite{metamorphic-01,metamorphic-02}. In addition, various types of assertions have been proposed, including dynamic assertions~\cite{li2020projection,dynamic-assertion-01,dynamic-assertion-02} and statistical assertions~\cite{statistical-assertion-01}, which open new opportunities and challenges for advancing the quality assurance of quantum programs.

\begin{figure*}[!tb]
\centering
\includegraphics[width=0.9\textwidth]{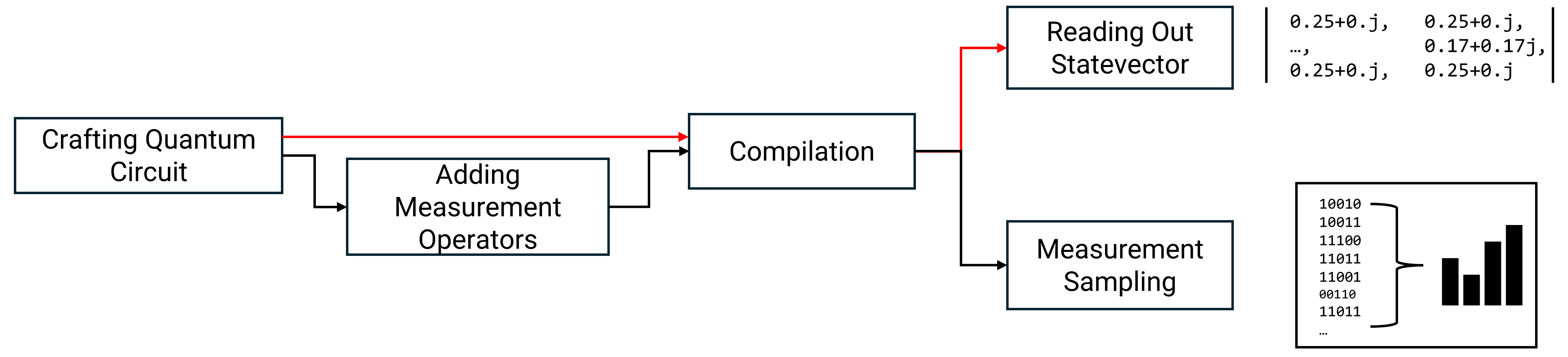}
\caption{The comparison of the process of reading out the statevector (in red arrow) and measurement sampling (in black arrow).}
\label{fig:process}
\end{figure*}

Our investigation of state-of-the-art quantum program testing approaches reveals a critical commonality: most validation methods rely on analyzing the outputs obtained from the measurement results, namely \emph{measurement-based validation}. In contrast, statevector-based validation methods are rarely adopted in existing approaches. The working processes of these two validation methods are illustrated in~\autoref{fig:process}. Unlike measurement-based validation, statevector-based validation does not require adding measurement operators to the circuit and directly extracts results without repeatedly sampling measurements.

Through our experiments and discussions with Qiskit developers, we identified a severe limitation of measurement-based validation in testing practice. Specifically, due to the probabilistic nature of quantum computing, the appearance of each output value in a single execution follows a particular probability distribution. Existing testing approaches extract measurement values and apply statistical analysis techniques, such as the Kolmogorov-Smirnov test~\cite{an1933sulla, smirnov1939estimation} and cross entropy~\cite{good1956some, shore2003axiomatic}, for output validation. However, determining the sufficient number of measurements to accurately capture the underlying probability distribution remains an inherently challenging task. Consequently, existing approaches often use ``undersampled'' distribution results for validation, particularly in differential testing, where distribution results are used to compare program behaviors. This raises fundamental questions: \emph{Are existing measurement-based validation methods sufficient for quantum program testing? Can statevector-based validation complement the shortcomings of existing approaches?}

To address these questions, we conduct a comprehensive survey of recent quantum program testing approaches to investigate how measurement-based validation participates in testing approaches. Specifically, we categorize existing testing approaches into distribution-level and output-value-level validation methods to answer our first research question (\textbf{RQ1}): \emph{In what ways do current testing approaches utilize measurement outcomes to assess program correctness?} Subsequently, we perform an in-depth comparison between statevector-based and measurement-based validation to reveal appropriate usage scenarios for both methods and discuss practical applications of statevector-based approaches, addressing our second research question (\textbf{RQ2}): \emph{What are the pros and cons of each validation method? How should these validation methods be appropriately applied in practice?} Based on our findings, we recommend using statevectors for complex validation tasks and measurement-based validation for simple validation requirements.

\noindent\textbf{Contributions.} Our main contributions are as follows:
\begin{enumerate}
    \item We conduct a comprehensive analysis of the validation methods employed in state-of-the-art quantum program testing approaches.
    \item We propose using statevectors as a complementary approach to measurement-based validation in quantum program testing.
    \item We provide a detailed comparison of both validation methods and provide practical guidelines for their appropriate application.
\end{enumerate}

\section{Background}
\label{sec:2 background}

In this section, we introduce the background and concepts related to this work.

\subsection{Qubit, Gates, Circuit, and Quantum Program}

A quantum bit (or \emph{qubit}) is the most basic unit in quantum computation. 
In quantum programs, the operations are represented by quantum logic gates. The gates provide various functions, including rotating a qubit to an arbitrary angle, assigning superposition to qubits, and creating controlled connections between qubits, among others. 
The quantum circuit is the most basic unit for grouping and managing qubits and quantum gates. A quantum circuit is a model for quantum computation~\cite{quantum-book}, in which the qubits, gates, and measurements are organized in a sequence. 
The quantum program combines the above quantum components and can further include the classical components. 
The development of modern quantum libraries enables users to manipulate quantum circuits using Python, thus facilitating the integration of quantum and classical components in quantum programs. 

\subsection{Statevector}

The statevector in a quantum program represents the complete quantum state of a quantum system as a complex-valued vector in the Hilbert space. For an n-qubit quantum system the statevector $\ket{\psi}$ is defined as:
\begin{equation*}
    \ket{\psi} = \sum_{\ket{x}\in B}{} A_{x} \ket{x}
\end{equation*}
where $A_{x}$ is the complex probability amplitudes and the normalization condition $\sum_{\ket{x}\in B}{} \lvert A_{x} \rvert^{2} = 1$ must be satisfied.

In quantum programs, the statevector is computationally represented as an array of complex numbers $2^n$, where each index of the array corresponds to a specific basis state $x$. The amplitude at each index determines the probability of measuring the system in that particular computational basis state, with the measurement probability given by $ \lvert A_{x} \rvert^{2}$.

The statevector of a quantum program is under the action of quantum gate operations as unitary transformations, which can be represented as $\ket{\psi} = U \ket{\psi'}$, where $U$ is a unitary operator preserving the normalization property. 

From a quantum program developer's perspective, the statevector serves as the fundamental data structure for quantum state representation in simulators and quantum virtual machines. The statevector enables precise tracking of quantum superposition and entanglement throughout program execution.

\subsection{Measuring the Output of a Quantum Program}

To read out the value of a qubit, a measurement operation can be added to the qubit at the end of the program. The measurement operation causes the quantum state to collapse. As explained previously, a qubit state $\ket{\psi}$ exists in a superposition of two orthogonal basis states $\ket{0}$ and $\ket{1}$. After measurement, the state collapses to a definite value of either 0 or 1, with probabilities determined by the amplitude of the superposition.

In the general case of an $n$-qubit program, we typically add $n$ measurement operations, one for each qubit, at the end of the program. After measurement, the results are read as a $n$-bit binary number. Due to the probabilistic nature of the output values, a single measurement is usually insufficient to draw meaningful conclusions. To address this issue, a solution is to repeat the measurement process, which is commonly referred to as repeated $m$ measurements (or samplings). According to the law of large numbers, with an increasing number of samplings, the frequency of observed values is likely to converge to the actual probability distribution. For a quantum program, if the number of samplings is sufficiently large, the distribution of output values should be close to the actual distribution. However, regardless of how well a sampling distribution approximates the actual distribution, it inherently introduces bias and overlooks critical information, thereby causing misjudgments in quantum program testing.

\section{Survey}

\begin{table}[tb]
  \centering
  \caption{A survey of quantum program testing literature. Distribution-level approaches treat measurement results as statistical distributions for analysis, whereas output-value-level approaches regard measurement results as plain output values.}
  \begin{minipage}{0.45\textwidth} 
  \end{minipage}
  \label{tbl:survey}
  \scriptsize
  \begin{tabular}{ccccc}
    \toprule
    Literature & Category & \makecell{Measurement\\-based} & \makecell{Distribution\\-level} & \makecell{Output-\\value-level}  \\
    \midrule
    \cite{search-based-01} & Search-based & \CIRCLE & \CIRCLE & \CIRCLE \\
    \cite{search-based-02} & Search-based & \CIRCLE & \CIRCLE & \CIRCLE \\
    \cite{combinatorial-01} & Combinatorial & \CIRCLE & \CIRCLE & \CIRCLE \\
    \cite{combinatorial-02} & Combinatorial & \CIRCLE & \CIRCLE & \CIRCLE \\
    \cite{metamorphic-01} & Metamorphic Testing & \CIRCLE & \CIRCLE & \CIRCLE \\
    \cite{metamorphic-02} & Metamorphic Testing & \CIRCLE & \CIRCLE & \Circle \\
    \cite{differential-testing-01} & Differential Testing & \CIRCLE & \CIRCLE & \Circle \\
    \cite{differential-testing-02} & Differential Testing & \CIRCLE & \CIRCLE & \Circle \\
    \cite{fuzzing-01} & Fuzzing & \CIRCLE & \Circle & \CIRCLE \\
    \cite{property-based-testing-01} & Property-based Testing & \CIRCLE & \Circle & \CIRCLE \\
    \cite{concolic-testing-01} & Concolic Testing & \CIRCLE & \CIRCLE & \Circle \\
    \cite{dynamic-assertion-01} & Dynamic Assertion & \CIRCLE & \Circle & \CIRCLE \\
    \cite{dynamic-assertion-02} & Dynamic Assertion & \CIRCLE & \Circle & \CIRCLE \\
    \cite{statistical-assertion-01} & Statistical Assertion & \CIRCLE & \Circle & \CIRCLE \\
    \cite{equivalence-check-01} & Equivalence Check & \Circle & \Circle & \Circle \\
    \cite{intergration-testing-01} & Integration Testing & \Circle & \Circle & \Circle \\
    \bottomrule
  \end{tabular}
\end{table}

To understand current research trends on validation methods in quantum program testing, we conduct a systematic survey of the recent literature, as illustrated in~\autoref{tbl:survey}. This table presents our comprehensive analysis of the quantum program testing literature, categorizing validation methods using three key dimensions: measurement-based testing, distribution-level validation, and output-value-level validation. Note that distribution-level and output-value-level validation represent two subcategories within measurement-based testing approaches. We employ different circle symbols with varying fill levels to indicate the degree of support each work provides across these dimensions. Specifically, $\CIRCLE$ means that the validation method is adopted, and $\Circle$ means not adopted. Our survey includes mainstream quantum program testing approaches, covering 11 distinct categories, including search-based testing, combinatorial testing, metamorphic testing, differential testing, fuzzing, property-based testing, concolic testing, dynamic assertion, statistical assertion, equivalence checking, and integration testing.

Our analysis reveals that most existing approaches adopt measurement-based methods, with only two exceptions~\cite{equivalence-check-01, intergration-testing-01}, which use unitary operators or statevectors for validation, thus avoiding measurement-based techniques. For the remaining approaches, they rely on measurement operations to obtain results for subsequent analysis. Based on how measurement results are utilized, we categorize these approaches into two groups: distribution-level and output-value-level validation. Distribution-level validation represents approaches that conduct validation based on the statistical distribution of repeated program executions, while output-value-level validation represents approaches that analyze individual measurement outcomes directly.

Distribution-level validation is used by 8 of the 15 approaches surveyed. Among these, five approaches support both the distribution-level and output-value-level analyzes, while the remaining three focus exclusively on distribution-level analyzes. Our investigation reveals that search-based approaches~\cite{search-based-01, search-based-02}, combinatorial approaches~\cite{combinatorial-01, combinatorial-02}, and a metamorphic approach~\cite{metamorphic-01} are designed to validate the existence of specific output values. In contrast, other approaches, such as the second metamorphic method~\cite{metamorphic-02}, differential testing~\cite{differential-testing-01,differential-testing-02}, and concolic testing~\cite{concolic-testing-01} target more complex validation tasks, including program behavior analysis and verification of output equivalence.

The validation of Output-values is adopted by 10 of the 15 approaches. Among these, five approaches support both distribution-level and output-value-level analysis, while the remaining 5 focus solely on output-value-level validation. Specifically, the fuzzing approach~\cite{fuzzing-01} uses measurements to assess the probability of occurrence for each output value. The property-based approach~\cite{property-based-testing-01} counts the occurrences of the output value for validation purposes. Dynamic assertion approaches~\cite{dynamic-assertion-01,dynamic-assertion-02} and statistical assertion methods~\cite{statistical-assertion-01} focus on detecting the appearance of unexpected output values.

\begin{tcolorbox}[size=title,rightrule=1mm, leftrule=1mm, toprule=0mm, bottomrule=0mm, arc=0pt,colback=gray!5,colframe=myblue,breakable]
{ \textbf{Answer to RQ1: } 
Most state-of-the-art testing approaches for quantum programs rely on measurement-based validation methods. Among these methods, distribution-level validation is used for complex validation tasks such astasks, such as comparing behaviors between two programs, while output-value-level validation is adopted for simpler tasks, like of specific output values.
}
\end{tcolorbox}
\section{Statevector vs. Measurement}

\subsection{Limitations of Measurement-Based Validation}

Measurement of a quantum program provides a direct method for obtaining program outputs for validation and behavioral assessment. Measurement is typically implemented by adding measurement gates to each qubit at the end of the quantum circuit. Since measurement gates always yield definite values from quantum circuits, the measurement process causes quantum state collapse, resulting in the loss of phase and amplitude information. After sampling (i.e., executing the quantum circuit), an output value is obtained. When multiple sampling repetitions are performed (e.g., 10,000 shots), the output values can be aggregated into a probability distribution. Existing testing approaches apply statistical methods to these distributions, interpreting divergences between programs as behavioral differences. For example, the widely used Kolmogorov-Smirnov test~\cite{an1933sulla} assesses distributional differences to detect anomalous program behavior. In contrast, the statevector provides a mathematically rigorous representation of quantum states. Unlike measurement, statevectors do not require additional circuit gates, and their extraction does not cause quantum state collapse. A workflow comparison is illustrated in~\autoref{fig:process}.

\begin{figure}[!tb]
\centering
\includegraphics[width=0.45\textwidth]{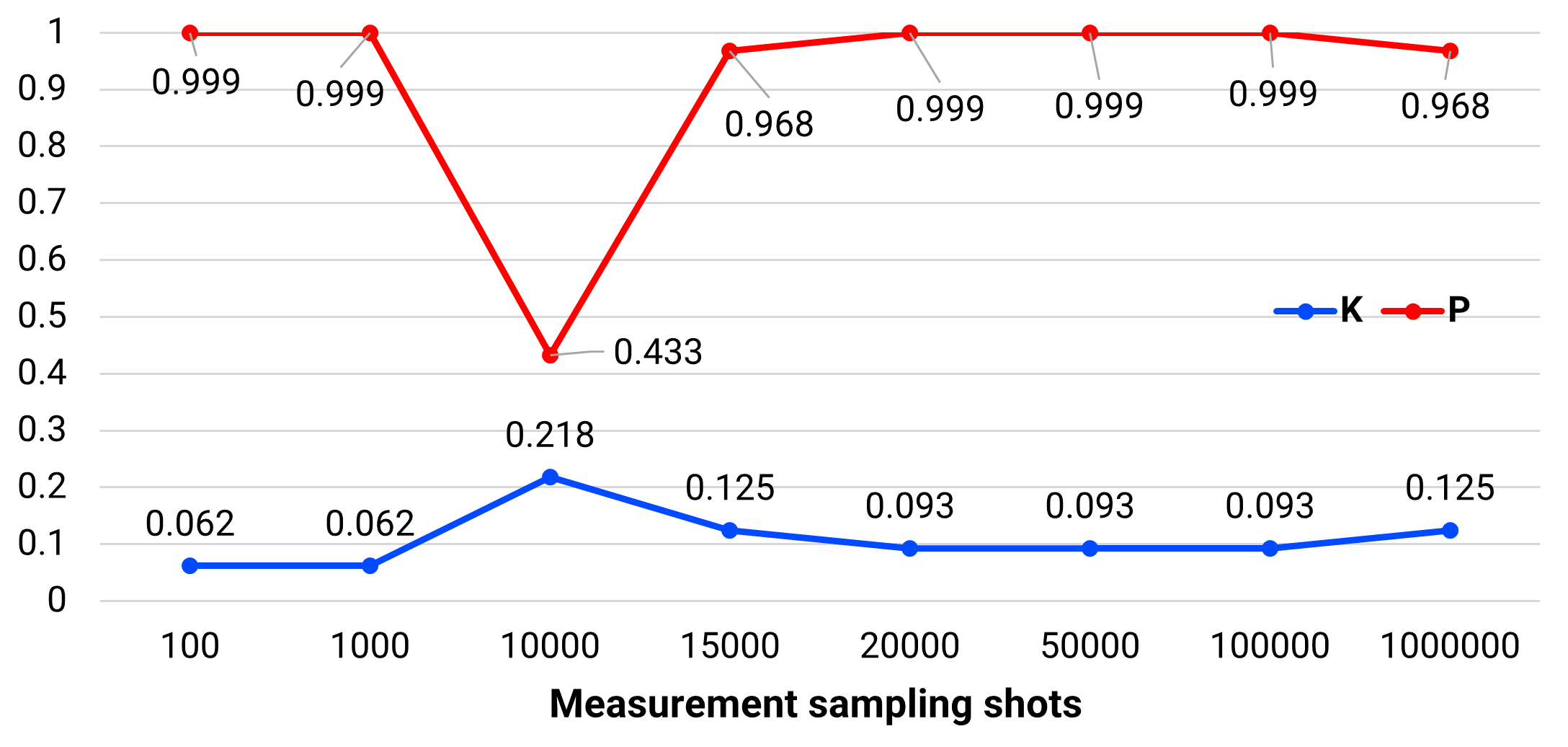}
\caption{The K and P values with the increasing measurement sampling shots.}
\label{fig:measure}
\end{figure}

We identify a major limitation in measurement-based output validation: measurement results exhibit inherent inconsistency. Since measurement results are derived from quantum circuit sampling, the observed distribution represents only an approximation of the true underlying distribution. A critical question in measurement-based testing is: How many sampling shots are sufficient to make the distribution representative of the true distribution? This question remains underexplored in the existing literature. Current approaches typically employ empirical shot numbers without rigorously assessing their sufficiency. Consequently, the inherent uncertainty in the measurements undermines the reliability of the downstream analyses.

Consider QDiff~\cite{differential-testing-01} as a representative example, where measurement results are used to determine the program's correctness. The approach executes programs repeatedly (approximately 10,000 times) to obtain sampling distributions, then applies the K-S test for statistical analysis. However, we observe that their results lack stability. When we randomly select a program and vary the sampling shot count, the results fluctuate significantly with changing sample sizes. As shown in~\autoref{fig:measure}, the key outputs of the Kolmogorov–Smirnov test, namely the K and P values, vary considerably as the number of sampling shots increases. Since QDiff's bug detection relies on pre-defined K-value thresholds, variations in sampling shots can significantly impact bug identification decisions.

Another critical limitation of measurement-based validation is scalability. In discussions with both freshmen and senior developers from Qiskit, Cirq, and Pytket, we found a consensus that measurement‐based approaches are confined to quantum program simulation. When a definite value is read out, the measurement leads to the quantum state collapse, which makes the quantum state unavailable for downstream applications. Furthermore, measurement-based validation methods are prone to introducing inherent noise from real quantum machines, which can lead to a misjudgment of the quantum program's behavior. These combined drawbacks, i.e., the nature of state collapse and the vulnerability to quantum noise, create significant obstacles for reliable quality assurance in quantum program development.

\subsection{The use of statevector-based validation}

Statevector-based validation offers several advantages over measurement-based approaches: 1) Stability. Unlike measurement-based methods that rely on statistical distribution comparisons, statevectors eliminate statistical uncertainties by directly representing quantum program states without probabilistic sampling. 2) Efficiency. Statevectors are obtained through single simulation runs, whereas measurement-based approaches require thousands of executions (typically 10,000+ shots) to construct distributions. 3) Scalability. Measurement-based methods require fixed simulator seeds to ensure reproducibility, which real quantum hardware cannot provide. Statevector-based validation avoids these constraints and can readily adapt to future quantum computing environments.

\begin{figure}[!tb]
\centering
\includegraphics[width=0.48\textwidth]{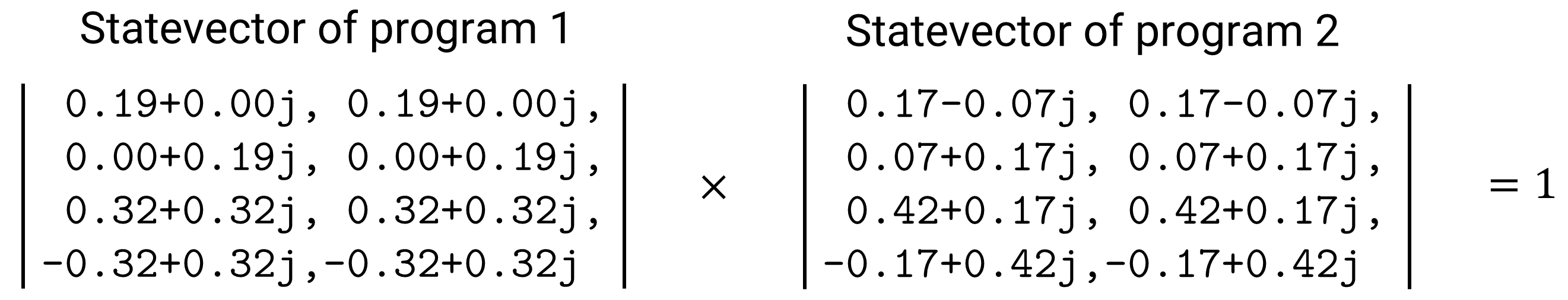}
\caption{An example of using the dot product for statevector-based validation.}
\label{fig:dotproduct}
\end{figure}

However, statevector-based validation should be carefully applied in practice.  After we discussed with the developers of Qiskit, we found that the global phase of a quantum circuit may affect the values of the statevector, and thus make the bit-to-bit comparison between two statevectors ineffective. To address this, we propose using the dot product of two statevectors to help decide whether they are different. Specifically, if the dot product of the two statevectors is neither 1.0 (which indicates that the statevectors are in the same direction) nor 0.0 (which means that the statevectors are in the opposite direction), such as 0.45, then the two statevectors can be considered as different. As shown in~\autoref{fig:dotproduct}, the statevector of program 1 and program 2 are different bit-wise due to the effect of the global phase. By calculating the dot product of the two statevectors, we can determine that they are the same.

\begin{tcolorbox}[size=title,rightrule=1mm, leftrule=1mm, toprule=0mm, bottomrule=0mm, arc=0pt,colback=gray!5,colframe=myblue,breakable]
{ \textbf{Answer to RQ2: } 
Measurement-based validation methods are suitable for simple assessment tasks, such as determining whether specific output values appear. For complex assessment tasks, such as comparing behaviors between two programs, measurement-based approaches show significant limitations, and using the calculation of statevectors can be a good alternative.
}
\end{tcolorbox}

Recent work by Miranskyy \textit{et al.}~\cite{miranskyy2025feasibility} further supports this view. Their quantitative evaluation shows that statevector-based validation achieves the best performance among several approaches, including inverse, swap, and statistical tests. In addition, they propose program reduction techniques that mitigate some of the scalability challenges associated with statevector simulation. These results strengthen the case for using statevectors in quantum program testing.

\section{Future Research}

Future research could focus on developing more robust and efficient measurement strategies to address the inherent challenges posed by current quantum validation methods. Traditional measurement is limited by statistical fluctuations, loss of phase information, and hardware-induced noise, especially in real quantum devices. By improving sampling techniques, incorporating error mitigation, and designing measurement-aware validation criteria, it may be possible to extract more meaningful insights from measurement results. Furthermore, adaptive measurement protocols and hybrid classical-quantum post-processing could further enhance the reliability and informativeness of measurement-based analyses.

In parallel, the statevector is expected to play an increasingly important role in future quantum program validation. The statevector allows for fine-grained comparison, equivalence checking, and structural validation of quantum programs, particularly in simulation settings. As tools for extracting and analyzing statevectors become more standardized and efficient, statevector-based validation will likely become a critical technique for validating program behavior in quantum program testing.

\section{Conclusion}

In this paper, we empirically analyze existing approaches for quantum program testing and find that measurement-based validation is primarily adopted in practice. Next, we present a short discussion to compare the measurement-based validation with the statevector. Finally, we summarize the limitations of measurement-based validation and the prospects for future use of the statevector.


\bibliographystyle{IEEEtran}
\bibliography{IEEEabrv,mybib}

\end{document}